\documentclass{emulateapj}

\usepackage{amssymb}
\usepackage{amsmath}
\usepackage{graphicx}
\usepackage{natbib}
\usepackage{txfonts}
\usepackage[colorlinks=true,citecolor=blue,linkcolor=blue]{hyperref}
\usepackage{microtype}
\usepackage{color}
\usepackage{subfigure} 
\bibpunct{(}{)}{;}{a}{}{,}

\def\mnras{MNRAS}
\def\apj{ApJ}
\def\apjl{ApJL}

\def\aap{A\&A}

\def\losalamos{1}
\def\icra{2}
\def\roma{3}
\def\rio{4}

\shorttitle{Angular momentum in BdHNe}
\shortauthors{}

\begin{document}

\title{Angular Momentum Role in the Hypercritical Accretion of Binary-Driven Hypernovae}

\author{L.~Becerra\altaffilmark{\roma,\icra},
				F.~Cipolletta\altaffilmark{\roma,\icra},
				Chris L.~Fryer\altaffilmark{\losalamos}, 
				Jorge A. Rueda\altaffilmark{\roma,\icra,\rio}, 
				Remo Ruffini\altaffilmark{\roma,\icra,\rio}}

\altaffiltext{\losalamos}{CCS-2, Los Alamos National Laboratory, Los Alamos, NM
87545}

\altaffiltext{\roma}{Dipartimento di Fisica and ICRA, 
                     Sapienza Universit\`a di Roma, 
                     P.le Aldo Moro 5, 
                     I--00185 Rome, 
                     Italy}
                     
\altaffiltext{\icra}{ICRANet, 
                     P.zza della Repubblica 10, 
                     I--65122 Pescara, 
                     Italy} 
                     
\altaffiltext{\rio}{ICRANet-Rio, 
                     Centro Brasileiro de Pesquisas F\'isicas, 
                     Rua Dr. Xavier Sigaud 150, 
                     22290--180 Rio de Janeiro, 
                     Brazil}                                    

\begin{abstract}
The induced gravitational collapse (IGC) paradigm explains a class of energetic, $E_{\rm iso}\gtrsim 10^{52}$~erg, long-duration gamma-ray bursts (GRBs) associated with Ic supernovae, recently named binary-driven hypernovae (BdHNe). The progenitor is a tight binary system formed of a carbon-oxygen (CO) core and a neutron star companion. The supernova ejecta of the exploding CO core trigger a hypercritical accretion process onto the neutron star, which reaches in a few seconds the  critical mass, and gravitationally collapses to a black hole emitting a GRB. In our previous simulations of this process we adopted a spherically symmetric approximation to compute the features of the hypercritical accretion process. We here present the first estimates of the angular momentum transported by the supernova ejecta, $L_{\rm acc}$, and perform numerical simulations of the angular momentum transfer to the neutron star during the hyperaccretion process in full general relativity. We show that the neutron star: i) reaches in a few seconds either mass-shedding limit or the secular axisymmetric instability depending on its initial mass; ii) reaches a maximum dimensionless angular momentum value, $[c J/(G M^2)]_{\rm max}\approx 0.7$; iii) can support less angular momentum than the one transported by supernova ejecta, $L_{\rm acc} > J_{\rm NS,max}$, hence there is an angular momentum excess which necessarily leads to jetted emission.
\end{abstract}


\maketitle

\section{Introduction}

We have introduced the concept of induced gravitational collapse \citep[IGC,][]{2008mgm..conf..368R,2012ApJ...758L...7R} and a family of systems that we have called binary-driven hypernovae \citep[BdHNe, see][and references therein]{2014A&A...565L..10R}, in order to explain the subfamily of gamma-ray bursts (GRBs) with energies $E_{\rm iso}\gtrsim 10^{52}$~erg associated with type Ic supernovae. Within this paradigm, the supernova explosion and the GRB occur in the following time sequence taking place in a binary system composed of a carbon-oxygen (CO) core and a neutron star (NS) companion: 1) explosion of the CO core; 2) hypercritical accretion onto the NS that reaches the critical mass; 3) NS gravitational collapse to a black hole; 4) emission of the GRB. This sequence occurs on short timescales of $\sim 100$ seconds in the source rest-frame, and it has been verified for several BdHNe with cosmological redshift $z \leq 1$ \citep{2013A&A...552L...5P}, all the way up to one of farthest sources, GRB 090423, at $z = 8.2$ \citep{2014A&A...569A..39R}.

The first theoretical treatment of the IGC process \citep{2012ApJ...758L...7R} was based on a simplified model of the binary parameters and on the
Bondi-Hoyle accretion formalism. More recently, \citet{2014ApJ...793L..36F} performed the first more realistic numerical simulations of the IGC by using more detailed supernova explosions coupled to hypercritical accretion models from previous simulations of supernova fallback \citep{1996ApJ...460..801F,2009ApJ...699..409F}. The core-collapse of the CO core producing the supernova Ic was simulated in order to calculate realistic profiles for the density and expanding velocity of the supernova ejecta. The hydrodynamic evolution of the material falling into the accretion region of the NS was there followed numerically up to the surface of the NS. The accretion in these systems can proceed at very high rates that exceed by several orders of magnitude the Eddington limit due to the fact that the photons are trapped in the accreting material and the accretion energy is lost through neutrino emission \citep[see][and references therein for additional details]{2014ApJ...793L..36F}.

In addition, \cite{FRR2015} have shown that BdHNe remain bound even if a large fraction of the binary initial total mass is lost in the explosion, exceeding the canonical 50\% limit of mass loss. Indeed, these binaries evolve through the supernova explosion very differntly than the compact binary progenitors studied for instance in population synthesis calculations due to the combined effects of: 1) the hypercritical accretion onto the NS companion; 2) the explosion timescale comparable with the orbital period; and 3) the bow shock created as the accreting NS plows through the supernova ejecta transfers angular momentum, acting as a break on the orbit.

In previous simulations, we adopted a spherically symmetric approximation of the hypercritical accretion process. However, the angular momentum that the supernova ejecta carry and eventually might transfer to the NS during the accretion process could play an important role in the evolution and fate of the system. In this work we give first, in section~\ref{sec:2}, an estimate of the angular momentum transported by the part of the supernova ejecta that enters into the gravitational capture region (Bondi-Hoyle surface) of the NS and compute the Bondi-Hoyle accretion rate. In section~\ref{sec:3} we show that the material entering into the Bondi-Hoyle region possesses sufficient angular momentum to circularize around the NS forming a disk-like structure. We then calculate in section~\ref{sec:4} the accretion of both matter and angular momentum onto the NS from the matter that have circularized into a disk. The accretion process is assumed to occur from an inner disk radius given by the most bound circular orbit around a rotating NS. We show that, depending upon the initial mass of the NS, the NS might reach either the mass-shedding limit or the secular axisymmetric instability in a short time. In section~\ref{sec:5} we evaluate, instead, the binary parameters for which the accretion onto the NS is not high enough to lead it to its collapse to a BH. The identification of such parameters defines a dichotomy in the final product of BdHNe which has been recently discussed by \cite{2015ApJ...798...10R}: Family-1 long GRBs in which the final system is a new NS binary, and Family-2 long GRBs in which a NS-BH binary is produced. Having introduced a typical CO core mass, the most important parameters that define such a dichotomy are the initial NS mass, the NS critical mass for the gravitational collapse to a BH, and the orbital period of the binary. We also discuss a possible evolutionary scenario leading to BdHN systems and compare and contrast our picture with existing binary evolution simulations in the literature. In the final discussion in section~\ref{sec:6} we summarize the results of this work and, in addition, show that the total angular momentum transported by the supernova ejecta is larger than the maximum angular momentum supported by a maximally rotating NS. Therefore, we advance the possibility that such an excess of angular momentum constitutes a channel for the formation of jetted emission during the hyperaccretion process of BdHNe leading to possible observable non-thermal high-energy emission.

\section{Angular momentum transported by the supernova ejecta}\label{sec:2}

\begin{figure}[!hbtp]
\includegraphics[width=\hsize,clip]{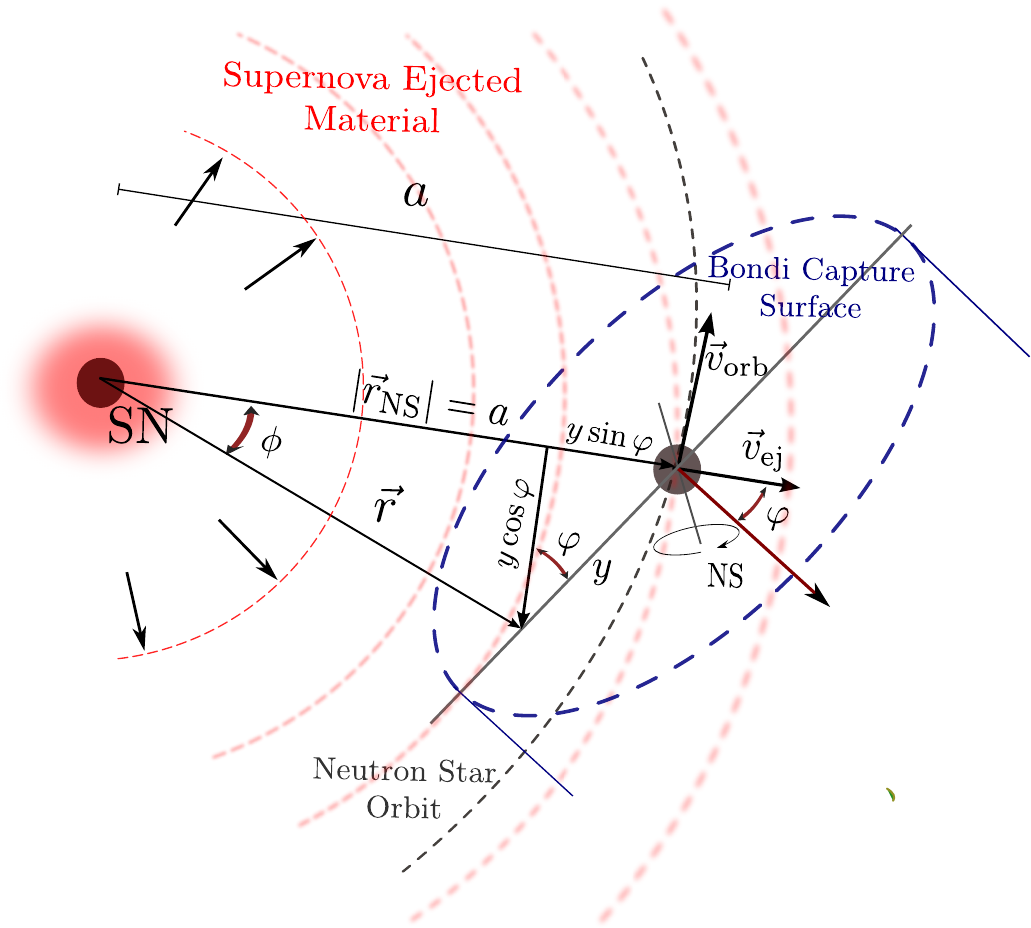}
\caption{Scheme of the IGC scenario: the CO core undergoes SN explosion, the NS accretes part of the SN ejecta and then reaches the critical mass for gravitational collapse to a black hole, with consequent emission of a GRB.}\label{fig:AccretionEsqueme}
\end{figure}

The accretion rate of the supernova ejecta onto the NS (see figure~\ref{fig:AccretionEsqueme}) can be estimated via the Bondi-Hoyle accretion formula \citep{1939PCPS...35..405H,1944MNRAS.104..273B,1952MNRAS.112..195B}:
\begin{equation}\label{eq:BondiMassRate_definition}
\begin{array}{rcl}
\dot{M}_B(t)&=&\pi\rho_{\rm ej}R_{{\rm cap}}^2\sqrt{v_{{\rm rel}}^2+c_{\rm s,ej}^2},
 \end{array}
\end{equation}
where $\rho_{\rm ej}$ is the density of the supernova ejecta, $R_{\rm cap}$ is the gravitational capture radius of the NS
\begin{equation}\label{eq:CaptureRadius}
R_{{\rm cap}}(t)=\frac{2 G M_{{\rm NS}}(t)}{v_{{\rm rel}}^2+c_{{\rm s,ej}}^2},
\end{equation}
with $G$ the gravitational constant, $M_{\rm NS}$ the mass of the NS, $c_{{\rm s,ej}}$ the sound speed of the supernova ejecta, and $v_{{\rm rel}}$ the ejecta velocity relative to the NS ($\vec{v}_{{\rm rel}}=\vec{v}_{{\rm orb}}-\vec{v}_{{\rm ej}}$, where $\vec{v}_{\rm orb}=\sqrt{G(M_{\rm core}+M_{\rm NS})/a}$ is the orbital velocity of the NS around the CO core, and $\vec{v}_{\rm ej}$ the velocity of the supernova ejecta).

Now, we are in position to give an estimate of the angular momentum transported by the supernova material to the NS during the IGC process. In doing so, we ``extrapolate'' the results of \cite{1976ShapiroLightman} and \cite{1981Wang} for the accretion process from stellar wind in a binary system. Due to the motion of the material and the orbital motion of NS, see figure~\ref{fig:AccretionEsqueme}, the material falls radially with a velocity $v_{\rm rel}$ making an angle $\varphi$ with respect to the line that joins the stars centers of the binary (so, $\sin\varphi=v_{\rm orb}/v_{\rm rel}$). Introducing Cartesian coordinates $(y,z)$ in the plane perpendicular to $\vec{v}_{\rm rel}(a)$, and putting the origin on the NS position, the angular momentum per unit time that crosses a surface element $dydz$ is:
\begin{equation}\label{eq:AngMom_definition}
d^2\dot{L}_{\rm acc}=\rho_{{\rm ej}}(y,z)v_{{\rm rel}}(y,z)^2 y\, dy dz.
\end{equation}
To first order in $y$, $\rho_{\rm ej}$ and $v_{\rm rel}$ can be written as:
\begin{equation}\label{eq:approx_var}
\rho_{\rm ej}(a)\simeq\rho_{\rm ej}(a)(1+\epsilon_\rho y) \;\; {\rm and} \;\; v_{\rm rel}(a)\simeq v_{\rm rel}(a)(1+\epsilon_{\nu} y),
\end{equation}
and equation (\ref{eq:AngMom_definition}) becomes
\begin{equation}
d^2\dot{L}_{\rm acc}=\rho_{{\rm ej}}(a)v_{{\rm rel}}^2(a)\left[y+(\epsilon_\rho+2\epsilon_{\nu})y^2\right]\,dydz.
\end{equation}
Integrating over the area of the circle delimited by the capture radius
\begin{equation}
y^2+z^2=R_{{\rm cap}}^2=\left(\frac{2GM_{{\rm NS}}}{v_{{\rm rel}}^2(a,t)}\right)^2\left(1-4\epsilon_\nu y\right),
\end{equation}
where we have applied $c_{\rm s,ej}\ll v_{\rm ej}$, we obtain the angular momentum per unit time of the ejecta material falling into the gravitational attraction region of the NS
\begin{equation}\label{eq:AngularMomentum_anal}
\dot{L}_{\rm acc}=\frac{\pi}{2}\,\left(\frac{1}{2}\epsilon_\rho-3\epsilon_{\nu}\right)\,\rho_{\rm ej}(a,t)v_{\rm rel}^2(a,t)R_{\rm cap}^4(a,t).
\end{equation}

Now, we have to evaluate the terms $\epsilon_\rho$ and $\epsilon_\nu$ of equation (\ref{eq:AngularMomentum_anal}). Following \cite{1987Anzeretal}, we start expanding $\rho_{\rm ej}(r)$ and $v_{\rm ej}$ in Taylor series around the binary separation distance, $r=a$:
\begin{equation}\label{eq:SNrho_expasion}
\rho_{{\rm ej}}(r,t)\approx\rho_{{\rm ej}}(a,t)\left(1+\frac{1}{\rho_{ej}(a,t)}\left.\frac{\partial \rho_{{\rm ej}}}{\partial r}\right|_{(a,t)}\delta r\right),
\end{equation}
\begin{equation}\label{eq:SNvel_expasion}
v_{{\rm ej}}(r,t)\approx v_{{\rm ej}}(a,t)\left(1+\frac{1}{v_{\rm ej}(a,t)}\,\left.\frac{\partial v_{\rm ej}}{\partial r}\right|_{(a,t)}\delta r\right),
\end{equation}
where we assumed $\delta r=|\vec{r}-\vec{r}_{\rm NS}|\ll 1$, keeping only the first order terms. For the supernova material, the continuity equation implies:
\begin{equation}
\frac{\partial \rho_{\rm ej}}{\partial t}=-\vec{\nabla}\cdot(\rho_{\rm ej}\,\vec{v}_{\rm ej}),
\end{equation}
and therefore we obtain
\begin{equation}
\epsilon_\rho=\left(\frac{\rho_{\rm ej}'}{\rho_{\rm ej}}\right)_{(a,t)}\sin\varphi=-\left(\frac{2}{r}+\frac{v_{\rm ej}'}{v_{\rm ej}}+\frac{1}{v_{\rm ej}}\,\frac{\dot{\rho}_{\rm ej}}{\rho_{\rm ej}}\right)_{(a,t)}\sin\varphi.
\end{equation}
On the other hand, defining $\hat{x}$ as a unit vector in the direction of $\vec{v}_{\rm rel}(a)$, the projection of $\vec{v}_{\rm rel}(r)$ on $\hat{x}$ is
\begin{equation}
\hat{x}\cdot \vec{v}_{\rm rel}(r)=v_{\rm ej}\cos(\phi+\varphi)-v_{\rm orb}\cos(\pi/2-\varphi)
\end{equation}
In the limit when $\delta r\ll 1$, also $\delta r\simeq -y\,\sin\varphi $ and $\sin\phi\simeq y/r \, \cos\phi$ (see figure \ref{fig:AccretionEsqueme}). Then, the last expression together with equation (\ref{eq:SNvel_expasion}) becomes
\begin{eqnarray}
\hat{x}\cdot \vec{v}_{\rm rel}(r)&\simeq& v_{\rm ej}(a)\,\cos\varphi+v_{\rm orb}\sin\varphi \nonumber \\
&-&\left(\frac{v_{\rm ej}}{r}+\frac{\partial v_{\rm ej}}{\partial r}\right)_{(a,t)}y\,\cos\varphi\,\sin\varphi .
\end{eqnarray}
We can write the relative velocity to the NS of the ejected material as
\begin{equation}\label{eq:approx_vrel}
v_{\rm rel}\simeq v_{\rm rel}(a,t)+\delta v_{\rm rel},
\end{equation}
where
\begin{equation*}
\delta v_{\rm rel}\simeq \hat{x}\cdot [\vec{v}_{\rm rel}(r)-\vec{v}_{\rm rel}(a)]=-\left(\frac{v_{\rm ej}}{r}+\frac{\partial v_{\rm ej}}{\partial r}\right)_{(a,t)}y\,\cos\varphi\,\sin\varphi.
\end{equation*}
Then, from a simple comparison of equations (\ref{eq:approx_var}) and (\ref{eq:approx_vrel}), we obtain:
\begin{equation}
\epsilon_\nu=-\left(\frac{v_{\rm ej}}{r}+\frac{\partial v_{\rm ej}}{\partial r}\right)_{(a,t)}\frac{{\rm cos}\, \varphi\,{\rm sin}\, \varphi}{v_{\rm rel}(a)}.
\end{equation}

In order to integrate equation (\ref{eq:BondiMassRate_definition}) and simulate the hypercritical accretion onto the NS, we need to implement a model for the supernova explosion from which we determine the velocity and the density of the ejecta near the capture region of the NS. We shall adopt for such a scope an homologous expansion of the supernova ejecta, i.e.:
\begin{equation}\label{eq:Sn_velocity}
v_{{\rm ej}}(r,t)=n\,\frac{r}{t}.
\end{equation}
Thus, the outermost layer of the ejecta, which we denote hereafter as $R_{{\rm star}}$, evolves as:
\begin{equation}\label{eq:radiusStar}
R_{{\rm star}}(t)=R_{0_{{\rm star}}}\left(\frac{t}{t_0}\right)^n,
\end{equation}
where $t_0>0$ is  the initial time of the accretion process and $n$ is the so-called {\it expansion parameter} whose value depends on the hydrodynamical evolution of the ejecta and the circumstellar material, i.e.: $n=1$ correspond to a free expansion, $n>1$ an accelerated expansion, and $n<1$ a decelerated one.

The condition of homologous expansion give us the density profile evolution \citep[see][for details]{1968Cox}:
\begin{equation}\label{eq:Sn_densityt}
\rho_{\rm ej}(X,t)=\rho_{\rm ej}(X,t_0)\frac{M_{\rm env}(t)}{M_{\rm env}(t_0)}\left(\frac{R_{0_{{\rm star}}}}{R_{\rm star}(t)}\right)^3\; \; X\equiv \frac{r}{R_{\rm star}},
\end{equation}
where $M_{\rm env}(t)$ is the mass expelled from the CO core in the supernova explosion, and hence available to be accreted by the NS, and $\rho_{\rm ej}(X,t_0)$ is the density profile of the outermost layers of the CO core (i.e., a pre-supernova profile). \cite{2014ApJ...793L..36F} considered the density profile for three different low-metallicity stars with initial zero-age main sequence (ZAMS) masses of $M_{\rm ZAMS} = 15$, $20$, and 30~$M_\odot$  using the Kepler
stellar evolution code \citep{2002WoosleyHegerWeaver}. The CO envelope of such pre-supernova configurations can be well approximated by a power law \citep[see figure 2 in][]{2014ApJ...793L..36F}:
\begin{equation}\label{eq:Sn_densityt0_1}
\rho_{\rm ej}(r,t_0)=\rho_{{\rm core}}\left(\cfrac{R_{\rm core}}{r}\right)^m, \quad {\rm for}\quad R_{{\rm core}}<r\leq R_{0_{\rm star}},
\end{equation}
where the parameters are shown in table \ref{tb:ProgenitorSN}.
\begin{table}[!hbtp]
\centering
\caption{Properties of the pre-supernova CO cores}\label{tb:ProgenitorSN}
\begin{tabular}{cccccc}
\hline \hline
Progenitor & $\rho_{\rm core}$& $R_{\rm core}$& $M_{\rm env}$&$R_{0\rm star}$&$m$ \\
$M_{\rm ZAMS}~(M_\odot)$ & (g~cm$^{-3}$) & (cm) & $(M_{\odot})$ & (cm) &\\\hline
$15$  & $3.31\times 10^8$& $5.01\times 10^7$ & $2.079 $  & $4.49\times 10^9$&$2.771$ \\
$20$   &$3.02\times 10^8$ &  $7.59\times 10^7$ & $3.89 $& $4.86\times 10^9$ &$2.946$ \\
$30$   & $3.08\times 10^8$& $8.32\times 10^7$ &  $7.94 $& $7.65\times 10^9$ &$2.801$ \\ \hline 
\end{tabular}
\tablecomments{The CO cores are obtained for the low-metallicity ZAMS progenitors with $M_{\rm ZAMS} = 15$, $20$, and 30~$M_\odot$ of \cite{2002WoosleyHegerWeaver}. The central iron core is assumed to have a mass $M_{\rm Fe}= 1.5~M_\odot$, which will be the mass of the new NS formed out of the supernova process.}
\end{table}

In the accretion process, the NS baryonic mass evolves as:
\begin{equation}
M_{b}(t)=M_{b}(t_0)+M_B(t),
\end{equation}
and, in general, its total mass which includes the gravitational binding energy, evolves as
\begin{equation}\label{eq:mgmb}
\dot{M}_{\rm NS}=\frac{\partial M_{\rm NS}}{\partial M_b}\dot{M}_{B}+\frac{\partial M_{\rm NS}}{\partial J_{\rm NS}}\dot{J}_{\rm NS}
\end{equation}
where $J_{\rm NS}$ is the NS total angular momentum. The relation between $M_b$ and $M_{\rm NS}$ for a rotating NS fully including the effects of rotation in general relativity, as well as other NS properties, are shown in appendix \ref{app:NSinfo}.

Taking into account the NS gravitational binding and considering the relations (\ref{eq:CaptureRadius}), (\ref{eq:Sn_velocity}), and (\ref{eq:Sn_densityt}), equation (\ref{eq:BondiMassRate_definition}) becomes:
\begin{equation}\label{eq:BondiMassRate}
\frac{\dot{\mu}_{B}}{(1-\chi\mu_{B})M_{\rm NS}(M_{b})^{2}}=\frac{t_{0}}{\tau_{B}}\,\frac{\tau^{n(m-3)}}{\hat{r}^{\, m}}\left[1+\eta\,\left(\frac{\hat{r}}{\tau}\right)^2\right]^{-3/2},
\end{equation}
with 
\begin{equation*}
\tau\equiv\frac{t}{t_0},\qquad \mu_{B}\equiv\frac{M_{B}(\tau)}{M_{\rm NS}(\tau=1)}, \qquad \hat{r}\equiv1-\frac{R_{B}}{a}.
\end{equation*}
 
The parameters $\chi$, $\eta$ and $\tau_{B}$ depend on the properties of the binary system before the SN explosion:
\begin{equation*}
\tau_{B}\equiv\frac{M_{\rm NS}(\tau=1) v_{\rm orb}^3}{4\pi G^2\rho_{\rm ej}(a,\tau=1)},\qquad \chi\equiv\frac{M_{\rm NS}(\tau=1)}{M_{\rm env}(\tau=1)},\qquad \eta\equiv\left(\frac{n}{t_{0}}\frac{a}{v_{\rm orb}}\right)^2.
\end{equation*}
%
For the homologous explosion model adopted to describe the expansion dynamics of the ejecta, the parameters $\epsilon_{\rho}$ and $\epsilon_{\nu}$, using equations (\ref{eq:Sn_densityt}) and (\ref{eq:Sn_densityt0_1}), are given by:
\begin{equation}
\epsilon_\rho(t)=-\frac{m}{a}\frac{v_{\rm orb}}{v_{\rm rel}(a,t)},\quad
\epsilon_\nu(t)=\frac{2}{a}\left(\frac{v_{\rm ej}}{v_{\rm rel}}\right)^2_{(a,t)}\frac{v_{\rm orb}}{v_{\rm rel}(a,t)}.
\end{equation}
Replacing the above equations in equation (\ref{eq:AngularMomentum_anal}), the angular momentum per unit time transported by the ejecta crossing the capture region is:
\begin{equation}\label{eq:AngMom_s}
\dot{L}_{\rm acc}=8\pi\,\rho_{\rm core}\left(\frac{R_{\rm core}}{a}\right)^m\frac{GM_{\rm NS}(t_0)a^2}{(1+q)^3} H(\tau),
\end{equation}
where
\begin{equation*}
H(\tau)=\tau^{n(m-3)}\left(\frac{M_{\rm NS}(\tau)}{M_{\rm NS}(\tau=1)}\right)^4(1-\chi\mu_B)\left(1+\frac{\eta}{\tau^2}\right)^{-7/2}\left(\frac{m}{2}+\frac{6\eta}{\tau^2+\eta}\right).
\end{equation*}
Thus, we can write the specific angular momentum:
\begin{equation}\label{eq:AngMom_g}
l_{\rm acc}=\frac{\dot{L}_{\rm acc}}{\dot{M}_B}=2\sqrt{\frac{GM_{\rm NS}(\tau=1)a}{(1+q)^3}} G(\tau),
\end{equation}
where
\begin{equation*}
G(\tau)=\hat{r}^m\left(\frac{M_{\rm NS}(\tau)}{M_{\rm NS}(\tau=1)}\right)^2\left(1+\frac{\eta}{\tau^2}\right)^{-3}\left(\frac{m}{2}+\frac{6\eta}{\tau^2+\eta}\right).
\end{equation*}

\begin{figure*}
\centering
\includegraphics[width=0.33\hsize,clip]{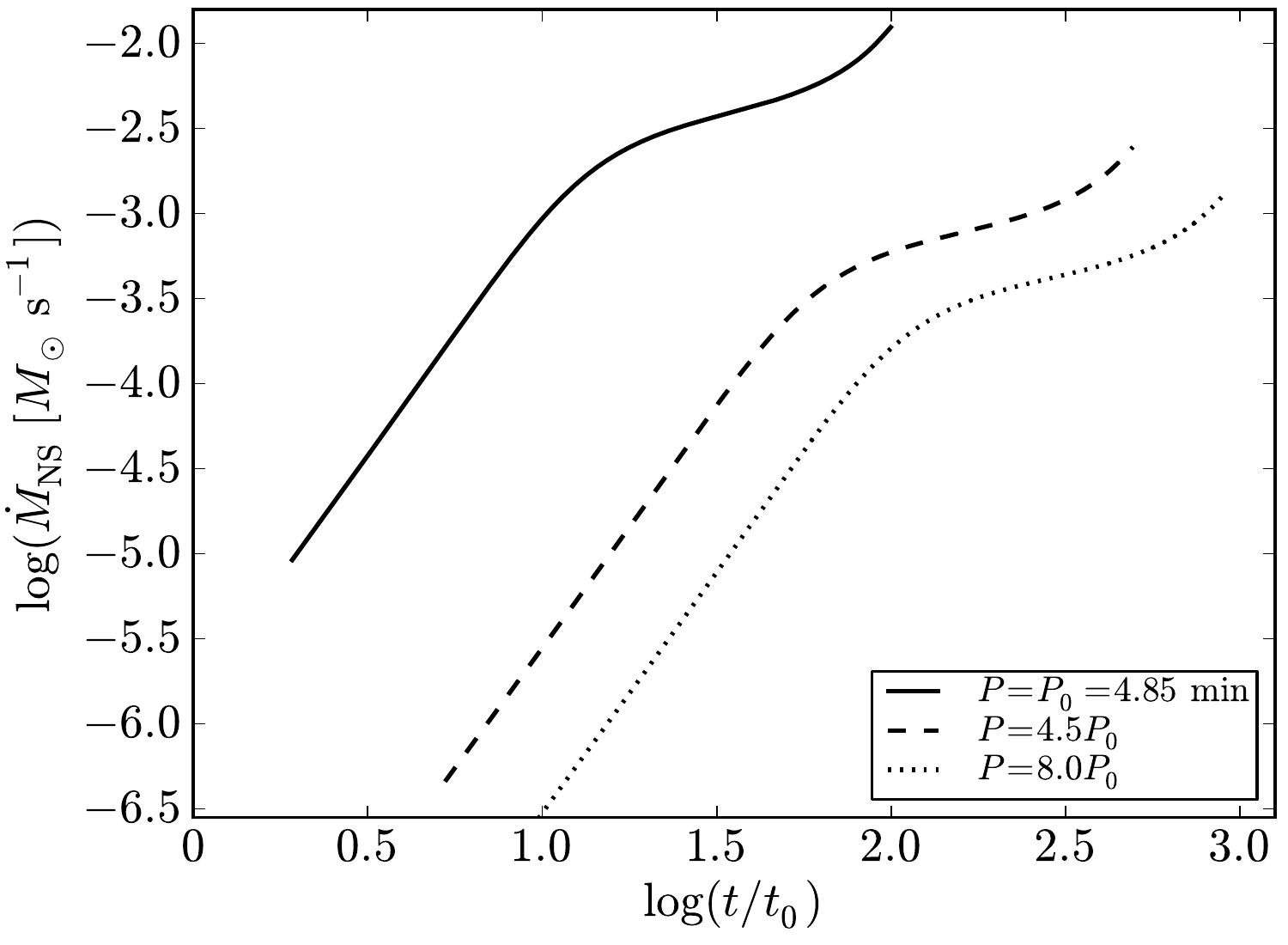}\includegraphics[width=0.33\hsize,clip]{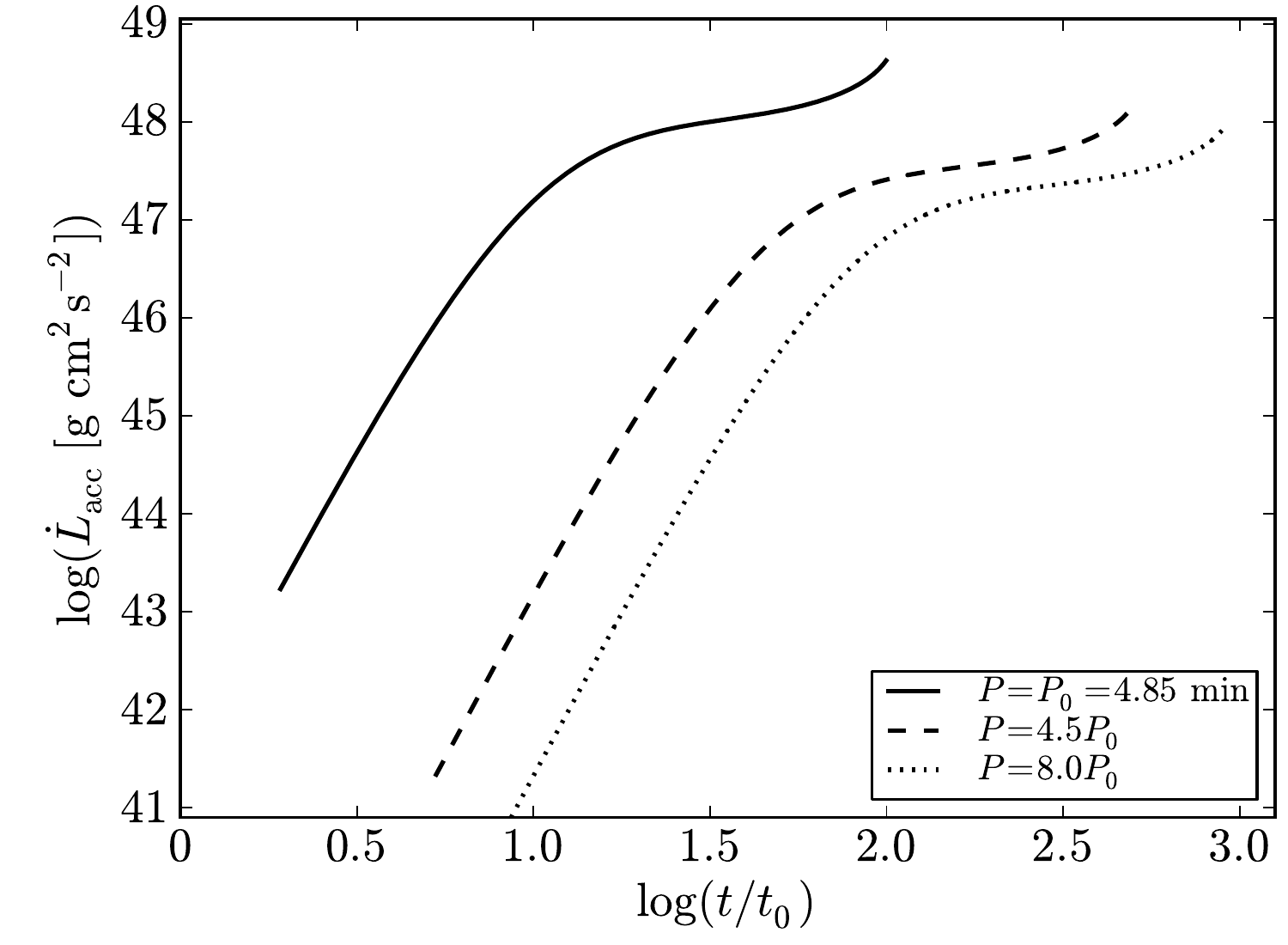}\includegraphics[width=0.33\hsize,clip]{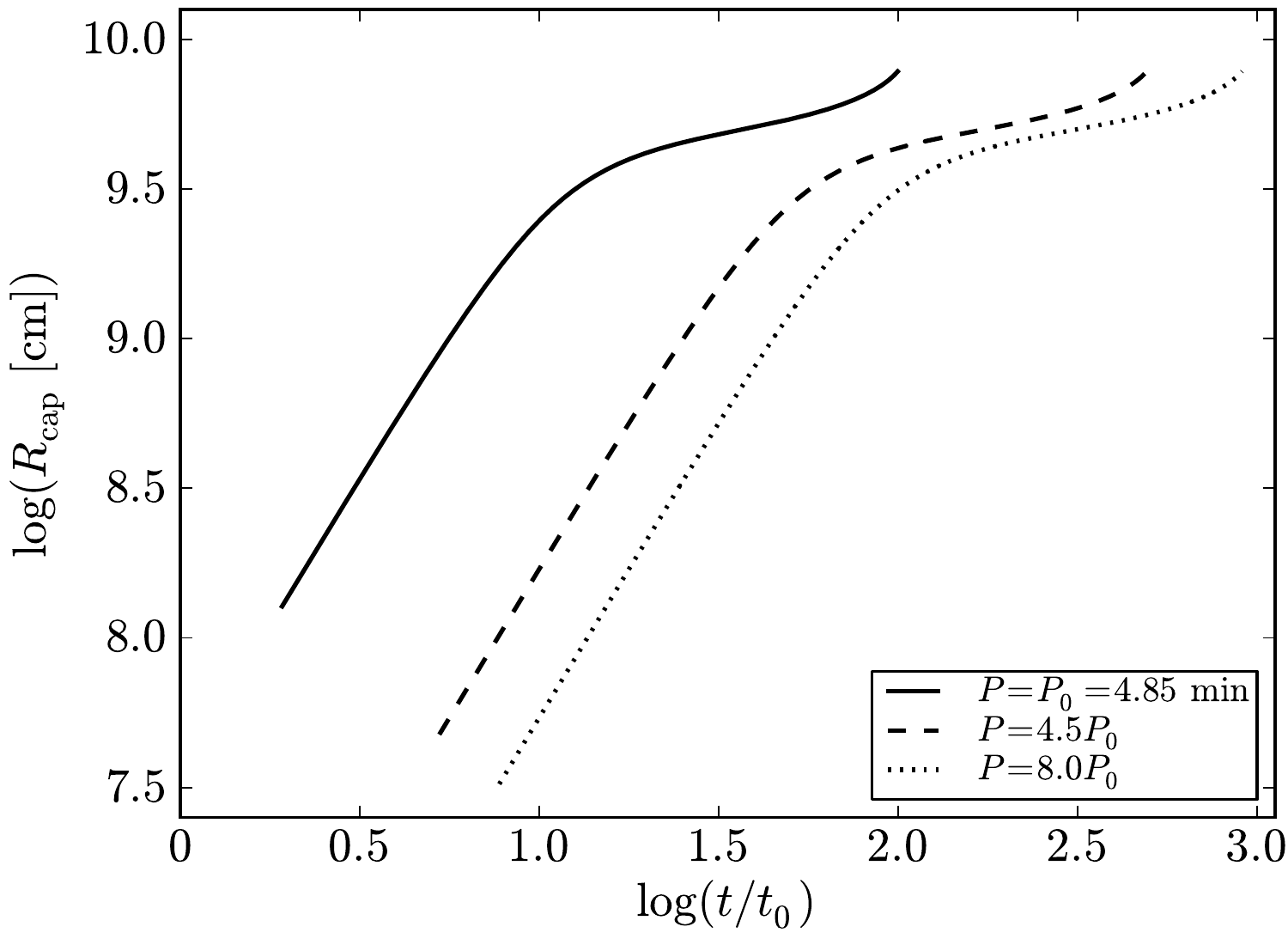}
\caption{Time evolution of the mass accretion rate (left panel, in units $M_\odot$~s$^{-1}$), the angular momentum per unit time transported by the supernova ejecta (central panel, in units g~cm$^2$~s$^{-2}$), and the Bondi-Hoyle capture radius of the NS (right panel, in units of $10^9$~cm).}\label{fig:AngularMomentum_IGC}
\end{figure*}

Figure \ref{fig:AngularMomentum_IGC} shows the time evolution of the Bondi-Hoyle accretion rate, obtained from the numerical integration of equation (\ref{eq:BondiMassRate}), and of the angular momentum transported by the ejecta, obtained from equation (\ref{eq:AngularMomentum_anal}). In these simulations we have adopted, for the sake of example, the $M_{\rm ZAMS}=30~M_\odot$ CO progenitor, an expansion parameter $n=1$, a supernova ejecta velocity $v_{0_{\rm star}}=2\times 10^9$~cm~s$^{-1}$, and an initial NS mass, $M_{\rm NS}(t_0)=2.0~M_\odot$. Following \cite{2014ApJ...793L..36F}, we adopt binary parameters such that there is no Roche lobe overflow prior to the supernova explosion. For the above CO core and NS parameters, such a condition implies a minimum orbital binary period $P_0=4.85$~min. 

\begin{figure*}[!hbtp]
\centering
\includegraphics[scale=0.35]{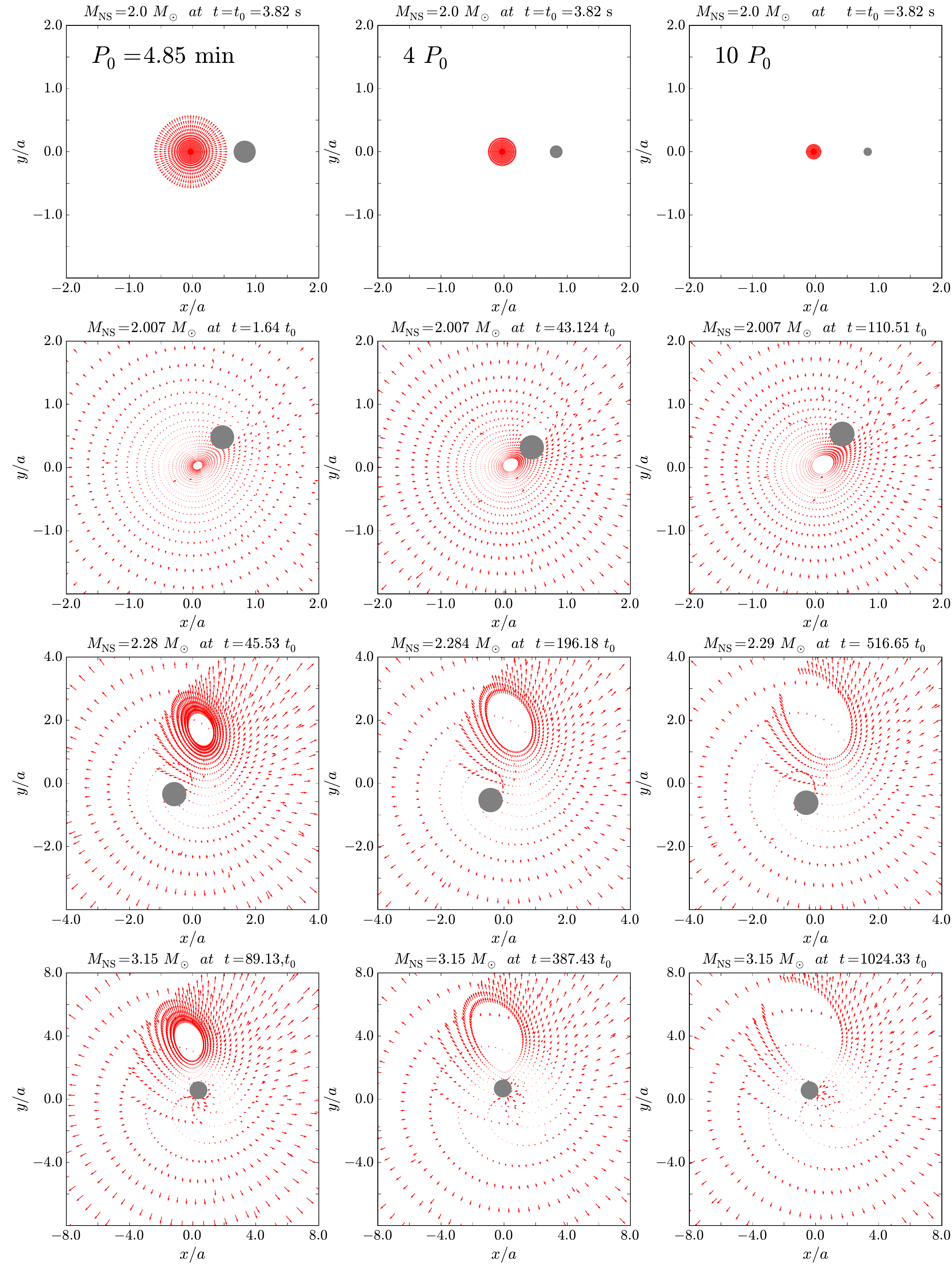}
\caption{Supernova ejecta velocity field at selected times of the accretion process onto the NS. This set of images show visibly the increase of mass of the NS as a function of time, and do manifestly evidence the physical reason why, as recently shown by \cite{FRR2015}, on the contrary to the canonical supernova explosion occurring in binary progenitors studied in population synthesis calculations, BdHNe remain bound even if a large fraction of the binary system's mass is lost in the explosion (i.e. well above the canonical 50\% limit of mass loss). The reason is first, the hypercritical accretion onto the NS companion alters both the mass and momentum of the binary; second, the explosion timescale is comparable with the orbital period, hence the mass ejection cannot be assumed to be instantaneous; and finally, the bow shock created as the accreting NS plows through the supernova ejecta transfers angular momentum, acting as a break on the orbit. In the simulation represented in these snapshots we have adopted the $M_{\rm ZAMS}=30~M_\odot$ CO progenitor (see table~\ref{tb:ProgenitorSN} for the corresponding pre-supernova CO core properties), an expansion parameter $n=1$, an ejecta outermost layer velocity $v_{0_{\rm star}}=2\times 10^9$~cm~s$^{-1}$, an initial NS mass, $M_{\rm NS}(t=t_0)=2.0~M_\odot$. We adopt binary parameters such that there is no Roche lobe overflow prior to the supernova explosion \citep{2014ApJ...793L..36F}, which imply for the present binary parameters a minimum orbital period of $P_0=4.85$~min. In the left, central and right columns of snapshots we show the results for binary periods $P=P_0$, $4 P_0$, and $10 P_0$, respectively. The Bondi-Hoyle surface, the filled gray circle, increases as the evolution continues mainly due to the increase of the NS mass. The x-y positions refer to the center-of-mass reference frame. The last image in each column corresponds to the instant when the NS reaches the critical mass value. For the initial conditions of these simulations, the NS ends its evolution at the mass-shedding limit with a maximum value of the angular momentum $J=6.14\times 10^{49}$~g~cm$^2$~s$^{-1}$ (or $j_{\rm NS}\approx 7$), and a corresponding critical mass of $M^{J\neq 0}_{\rm crit}=3.15~M_\odot$ (see, also, figures~\ref{fig:NSa} and \ref{fig:NSb}).}\label{fig:vfield}
\end{figure*}

In order to visualize the dynamics of the process, we show in figure \ref{fig:vfield} the velocity field of the supernova ejecta at selected times of the accretion process onto the NS. To produce this figure, we compute the SN velocity field considering just the effect of the NS gravitational attraction on the SN ejecta  taking into account the changes in the NS position, $\vec{r}_{\rm NS}(t)$, due by its own orbital motion and the evolution of the NS mass, $M_{\rm NS}(t)$, estimated by the Bondi accretion formalism. The SN matter is seeing as a set of point-like particles, so the trajectory of each particle was followed by solving the Newtonian equation of motion:
\begin{equation}
\frac{d^2 \vec{r}_{\rm sn}(t)}{dt^2}=-GM_{\rm NS}(t) \frac{\vec{r}_{\rm sn}(t)-\vec{r}_{\rm NS}(t)}{|\vec{r}_{\rm sn}(t)-\vec{r}_{\rm NS}(t)|^3},
\end{equation}
setting as reference frame the initial center of mass of the binary system. The simulation goes from time $t=t_0$ until the collapse of the NS. The initial conditions for each SN particle come from the homologous velocity distribution, assuming a free expansion.

\section{Circularization of the Supernova Ejecta Around the Neutron Star}\label{sec:3}

We turn now to estimate if the supernova ejecta possess enough angular momentum to circularize around the NS before being accreted by it. Since initially the NS is slowly rotating or non-rotating, we can describe the exterior spacetime of the NS before the accretion process by the Schwarzschild metric. A test-mass particle circular orbit of radius $r_{\rm st}$ possesses in this metric a specific angular momentum given by:
\begin{equation}
l_{\rm st}=c\sqrt{\frac{GM_{\rm NS}r_{\rm st}}{c^2}}\left(1-\frac{3GM_{\rm NS}}{c^2r_{\rm st}}\right)^{-1/2}.
\end{equation}
Assuming that there are no angular momentum losses once the ejected material enters the NS capture region, hence $l_{\rm st} = l_{\rm acc}=\dot{L}_{\rm acc}/\dot{M}_B$, the material circularizes around the NS at the radii:
\begin{equation}
r_{\rm st}=\frac{1}{2}\left[\frac{l_{\rm acc}^2}{GM_{\rm NS}}+\sqrt{\left(\frac{ l_{\rm acc}^2}{G M_{\rm NS}}\right)^2-12\left(\frac{l_{\rm acc}}{c}\right)^2}\right] .
\end{equation}
y|circular orbit around a non-rotating NS is located at a distance $r_{\rm mb}=6 G M_{\rm NS}/c^2$ with an angular momentum per unit mass $l_{\rm mb}=2\sqrt{3} G M_{\rm NS}/c$. The most bound circular orbit, $r_{\rm mb}$, is located outside the NS surface for masses larger than $1.57~M_\odot$, $1.61~M_\odot$, and $1.68~M_\odot$ for the GM1, TM1, and NL3 EOS, respectively \cite{2015Cippollettainprep}. It is easy to check (see figure~\ref{fig:AngularMomentum_IGC}) that the supernova ejected material has an angular momentum larger than this value, and therefore the ejecta entering into the NS capture region will necessarily circularize at radii $r_{\rm st}> r_{\rm mb}$. We have obtained from our simulations, $r_{\rm st}/r_{\rm mb}\sim 10$--$10^3$.

Even if the supernova ejecta possess enough angular momentum to form a disc around the NS, which prevent it to fall rapidly onto the NS, the viscous forces (and other angular momentum losses) might allow the material to be accreted by the NS when it arrives to the inner boundary of the disc. In the $\alpha$-disc formalism, the kinetic viscosity is $\nu=\eta/\rho=\alpha c_{s,\rm disk} H$, where $\eta$ is the dynamical viscosity, $H$ the disk scale height, and $c_{s,\rm disk}$ the sound velocity of the disk. Following \cite{1993Chevalier}, the infall time in a disk at radius $r$ is:
\begin{equation}\label{eq:TimeInfall}
t_{{\rm fall}}\sim \frac{r_{{\rm st}}^2}{\alpha c_{{\rm s,disk}}H}\sim \, \frac{r_{\rm st}^{3/2}}{\alpha\sqrt{GM_{\rm NS}}}\sqrt{1-\frac{2GM_{\rm NS}}{c^2r_{\rm st}}},
\end{equation}
where it is assumed that $H\sim r_{\rm st}$ (thick disk) and $c_{s,\rm disk}$ of the order of orbital velocity seen by an observer corotating with the particle. Finally, $\alpha\sim 0.01$--$0.1$ is dimensionless and measures the viscous stress. In our simulations we have obtained falling times $t_{\rm fall}/\Delta_{\rm acc}\sim 10^{-3}$, where $\Delta_{\rm acc}$ is the characteristic accretion time. Therefore, the supernova material can be accreted by the NS in a short time interval without introducing any significant delay.

With even a mild amount of angular momentum, this accretion drives a strong outflow along the axis of angular momentum \citep{2006ApJ...646L.131F,2009ApJ...699..409F}, ejecting up to 25\% of the infalling material, removing much of the angular momentum.  The ejecta may undergo rapid neutron capture producing r-process elements \citep{2006ApJ...646L.131F}.  As the angular momentum increases, we expect the outflows to become more collimated, which might very well lead to the long sought explanation of the high-energy power-law MeV emission observed in the Episode 1 of BdHNe (see, e.g., sections 5 and 7 in \citealp{2012A&A...543A..10I}; sections 3.2 and 3.3 in \citealp{2013arXiv1311.7432R,2015ARrees}).
Much more work is needed to determine if there are any observation implications of these expelled material.

\section{Spin-Up of the Neutron Star and Reaching of the Instability Region}\label{sec:4}

Our first estimate of the angular momentum transported by the supernova ejecta has shown that the material have enough angular momentum to circularize around the NS for a short time and form a kind of thick disc. The viscous force torques in the disc (and other possible losses) allow a sufficient angular momentum loss until the material arrives to the inner boundary of the disc, $R_{\rm in}$, then falling into the NS surface. Thus, from angular momentum conservation, the evolution of the NS angular momentum, $J_{\rm NS}$, will be given by:
\begin{equation}\label{eq:AngMomNSevolution}
\frac{d J}{dt}=\xi \, l(R_{\rm in})\frac{dM_{\rm B}}{dt} \,
\end{equation}
where $\xi$ is an efficiency angular momentum transfer parameter. In order to follow the NS spin-up during the accretion process, this equation have to be integrated simultaneously with equation (\ref{eq:BondiMassRate}). In equation (\ref{eq:AngMomNSevolution}) the value of $l(R_{\rm in})$ corresponds to the angular momentum of the mostly bound circular orbit, which for the axially symmetric exterior spacetime around the rotating NS can be written as \cite{2015Cippollettainprep}:
\begin{equation}\label{eq:lisco}
l_{\rm mb}=\frac{G M_{\rm NS}}{c}\left[ 3.464 - 0.37\left(\frac{j_{\rm NS}}{M_{\rm NS}}\right)^{0.85}\right]
\end{equation}

\begin{figure}[!hbtp]
\includegraphics[width=\hsize,clip]{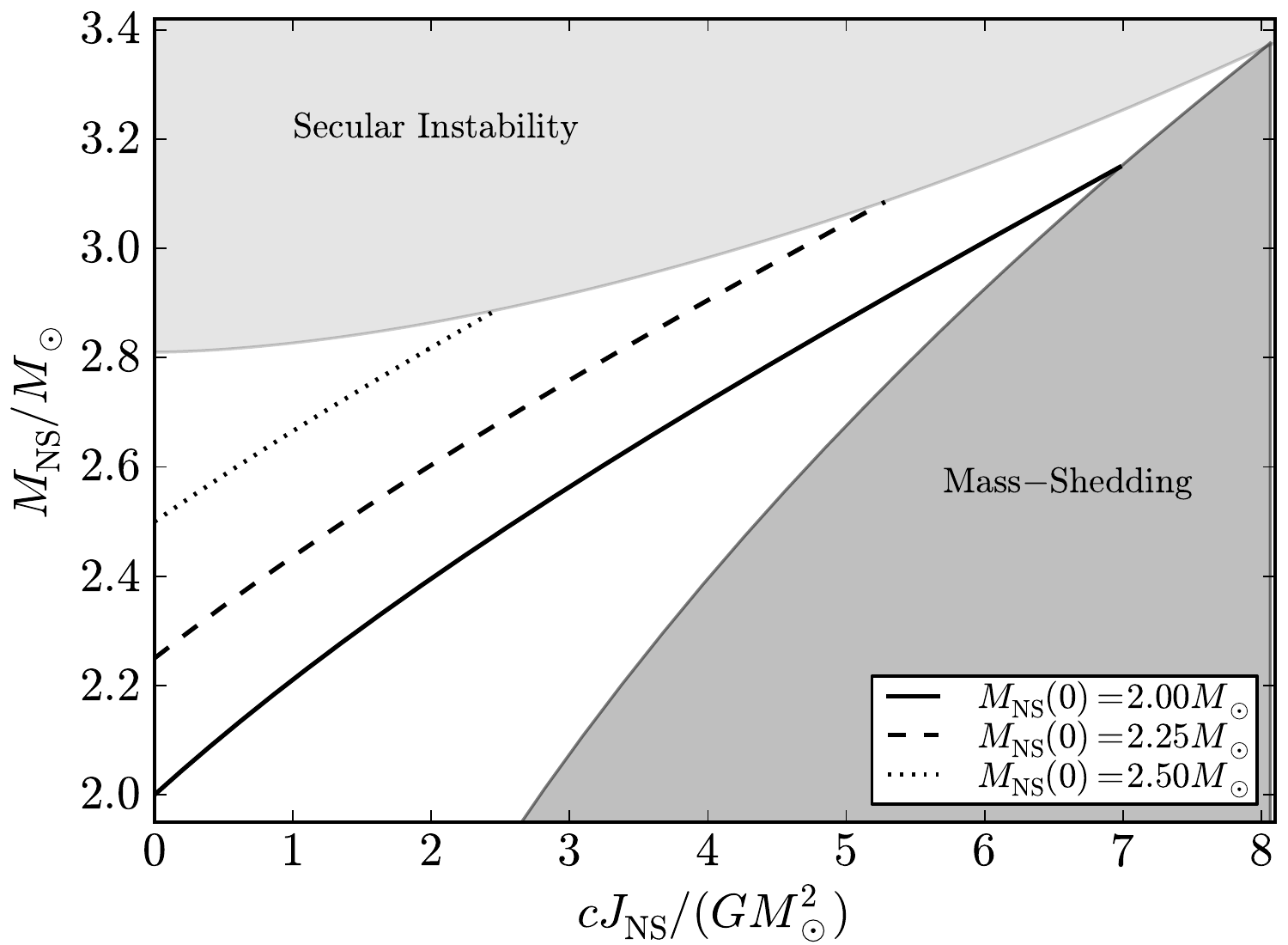}
\caption{Neutron star mass as a function of its angular momentum gain for a selected efficient parameter $\xi=0.5$. Three evolutionary paths of the NS are shown, each starting from a different initial NS mass $M_{\rm NS}(t=0)= 2.0$, $2.25$, and $2.5~M_\odot$, and without loss of generality we have adopted the NL3 nuclear EOS.}\label{fig:NSa}
\end{figure}

In figure \ref{fig:NSa}, we show the evolution of the NS mass as a function of its angular momentum during the accretion process for a selected efficiency parameter for the NS angular momentum rate, $\xi=0.5$, and for selected values of the initial NS mass, $M_{\rm NS}(t=0)= 2.0$, $2.25$, and $2.5~M_\odot$. We see how the NS starting with $M_{\rm NS}(t=0)=2.0~M_\odot$ reaches the mass-shedding limit while, for higher initial masses, the NS ends at the secular axisymmetric instability region.

\begin{figure}[!hbtp]
\includegraphics[width=\hsize,clip]{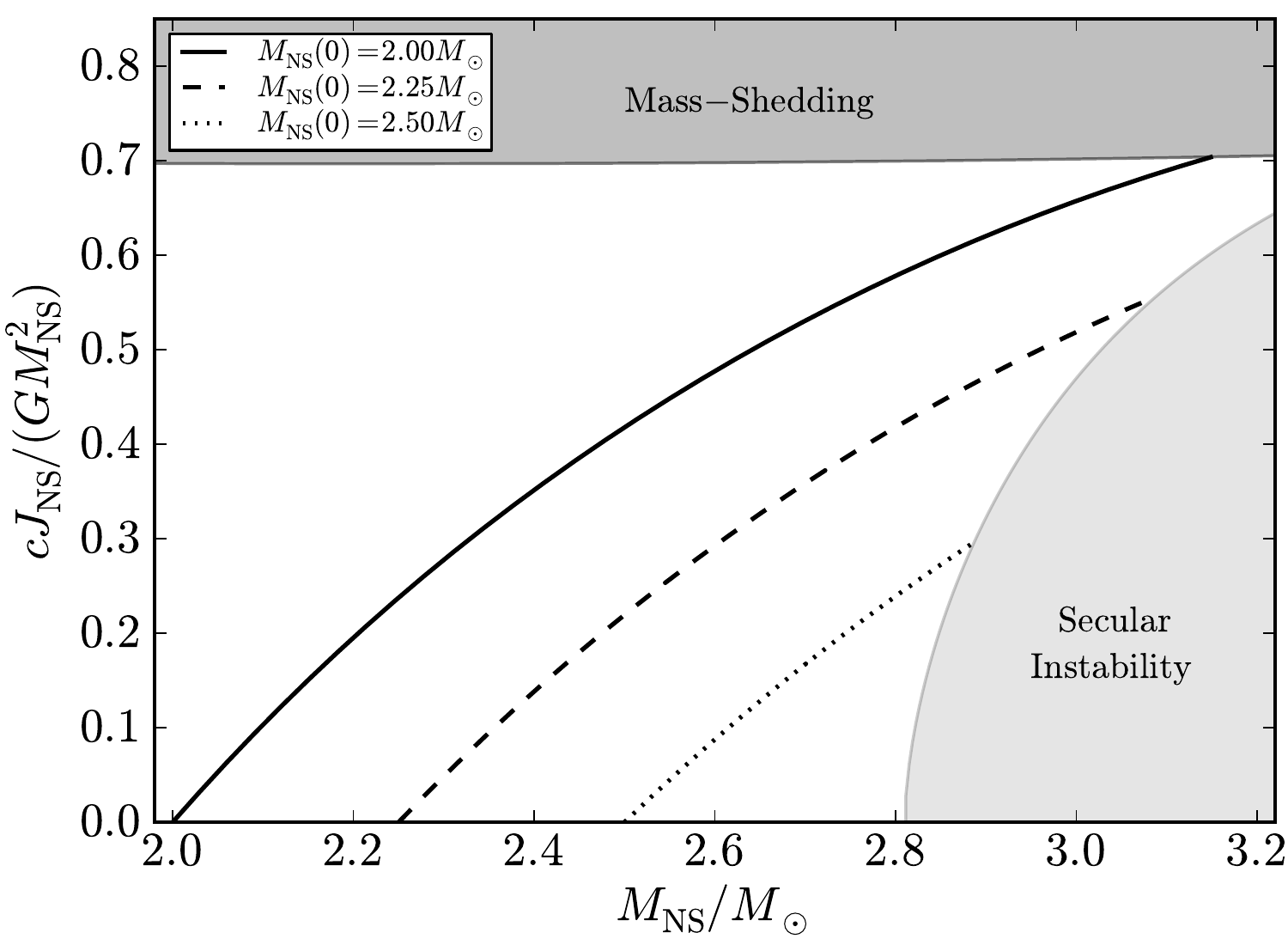}
\caption{Dimensionless angular momentum gain of the NS as a function of the NS mass for a selected efficiency parameter $\xi=0.5$. Three evolutionary paths of the NS are shown, each starting from a different initial NS mass $M_{\rm NS}(t=0)= 2.0$, $2.25$, and $2.5~M_\odot$, and without loss of generality we have adopted the NL3 nuclear EOS.}\label{fig:NSb}
\end{figure}

Correspondingly, we show in figure \ref{fig:NSb} the time evolution of the dimensionless angular momentum of the NS, $c J_{\rm NS}/(G M_{\rm NS}^2)$, as a function of the NS mass, in the case of an efficiency of angular momentum transfer to the NS $\xi=0.5$. In line with the above result of figure~\ref{fig:NSa}, we see here that only the NS that ends at the mass-shedding limit, i.e the one with an initial mass $M_{\rm NS}(t=0)=2.0~M_\odot$, reaches the maximum possible value of the dimensionless angular momentum, $c J_{\rm NS}/(G M_{\rm NS}^2)\approx 0.7$. The NSs with higher initial masses becomes secularly unstable with lower angular momentum.

\section{Family-1 and Family-2 Long GRBs and Evolutionary Scenario}\label{sec:5}

We have until now evidenced the binary parameters for which a new NS-BH binary system is produced out of the BdHNe. There exist limiting binary parameters leading to lower accretion rates onto the NS and to a total accreted matter not sufficient to bring the NS to the gravitational collapse to a BH, namely not sufficient to reach the NS critical mass. The identification of such limiting parameters introduce a dichotomy in the final product of BdHNe, with consequent different observational signatures, which have been recently discussed by \cite{2015ApJ...798...10R}: Family-1 long GRBs in which the NS does not collapse to a BH, and the final system being a new NS binary; and Family-2 long GRBs in which the NS collapses to a BH, leading to a new NS-BH binary.

\begin{figure}[!hbtp]
\includegraphics[width=\hsize,clip]{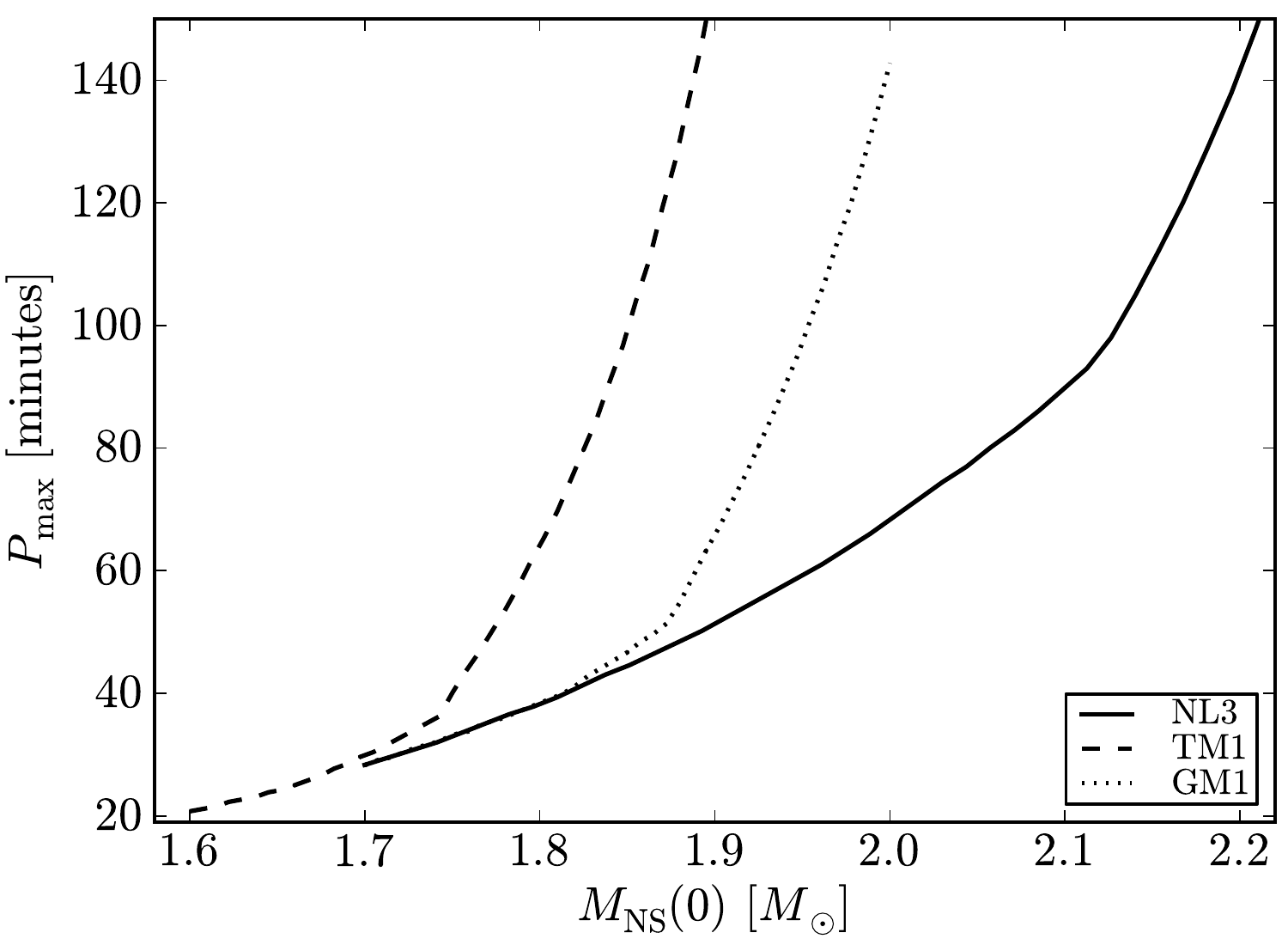}
\caption{Maximum binary period for which the NS with initial mass $M_{\rm NS}(0)$ collapses to a BH, for three selected EOS.}\label{fig:Pmax}
\end{figure}

Since we expect the accretion rate to become lower for wider (longer binary period) binaries, there must be a maximum binary period (of course having fixed the other system parameters, i.e. initial NS mass, CO core mass, and supernova velocity), which defines the aforementioned families dichotomy. Therefore, we performed simulations increasing the binary period, starting from the minimum value $P_0$, for given $M_{\rm NS}(0)$, $M_{\rm CO}$, and 
$v_{0_{\rm star}}$. Figure~\ref{fig:Pmax} shows the results of the maximum binary period $P_{\rm max}$ for which we obtained collapse to a BH of the NS companion, as a function of its initial mass, $M_{\rm NS}(0)$, keeping the other binary parameters fixed. The results are shown for three selected nuclear EOS of the NS (NL3, TM1 and GM1). In particular, for the system we have used as an example in the above figures, namely $M_{\rm NS}(0)=2~M_\odot$, $M_{\rm CO}=9.4~M_\odot$ ($M_{\rm ZAMS}=30~M_\odot$, see table~\ref{tb:ProgenitorSN}), and $v_{0_{\rm star}}=2\times 10^9$~cm~s$^{-1}$, we obtained $P_{\rm max}\approx 73$~min, for the NL3 EOS. Since the two other EOS (TM1 and GM1) lead to lower values of the critical NS mass with respect to the NL3 EOS (see table~\ref{tb:StaticRotatingNS} and \cite{2015Cippolletta}), the maximum orbital period to have BH formation is longer for the same initial NS mass.

With the knowledge of the proper set of parameters for which the gravitational collapse of the NS to a BH is induced by accretion, and consequently, also the parameters for which such a process does not occur, it becomes appropriate to discuss the possible progenitors of such binaries. A possible evolutionary scenario was discussed by \cite{2012ApJ...758L...7R}, taking advantage of the following facts: 1) a viable progenitor for BdHN systems is represented by X-ray binaries such as Cen X-3 and Her X-1 \citep{1972ApJ...172L..79S,1972ApJ...174L..27W,1972ApJ...174L.143T,1973ApJ...180L..15L,1975ASSL...48.....G,2011ApJ...730...25R}; 2) evolution sequences for X-ray binaries, and evolution scenarios leading to systems in which two SN events occur during their life, had been already envisaged \citep[see, e.g.,][]{1988PhR...163...13N,1994ApJ...437L.115I}. Thus, BdHNe could form following the evolution  of an initial binary system composed of two main-sequence stars, $M_1$ and $M_2$, with a mass ratio $M_2/M_1\gtrsim 0.4$. The star 1 is likely $M_1 \gtrsim 11~M_\odot$, leaving a NS through a first core-collapse event. The star 2, now with $M_2\gtrsim 11~M_\odot$ after some almost conservative mass transfer, evolves filling its Roche lobe. It then starts a spiral-in of the NS into the envelope of the star 2, i.e. a common-envelope phase occurs. If the binary system survives to this common-envelope phase, namely it does not merge, it will be composed of a helium star and a NS in tight orbit. The helium star then expands filling its Roche lobe and a non-conservative mass transfer to the NS, takes place. After loosing its helium envelope, the above scenario naturally leads to a binary system composed of a CO star and a massive NS whose fate has been discussed in the present work.

\begin{figure}[!hbtp]
\includegraphics[width=\hsize,clip]{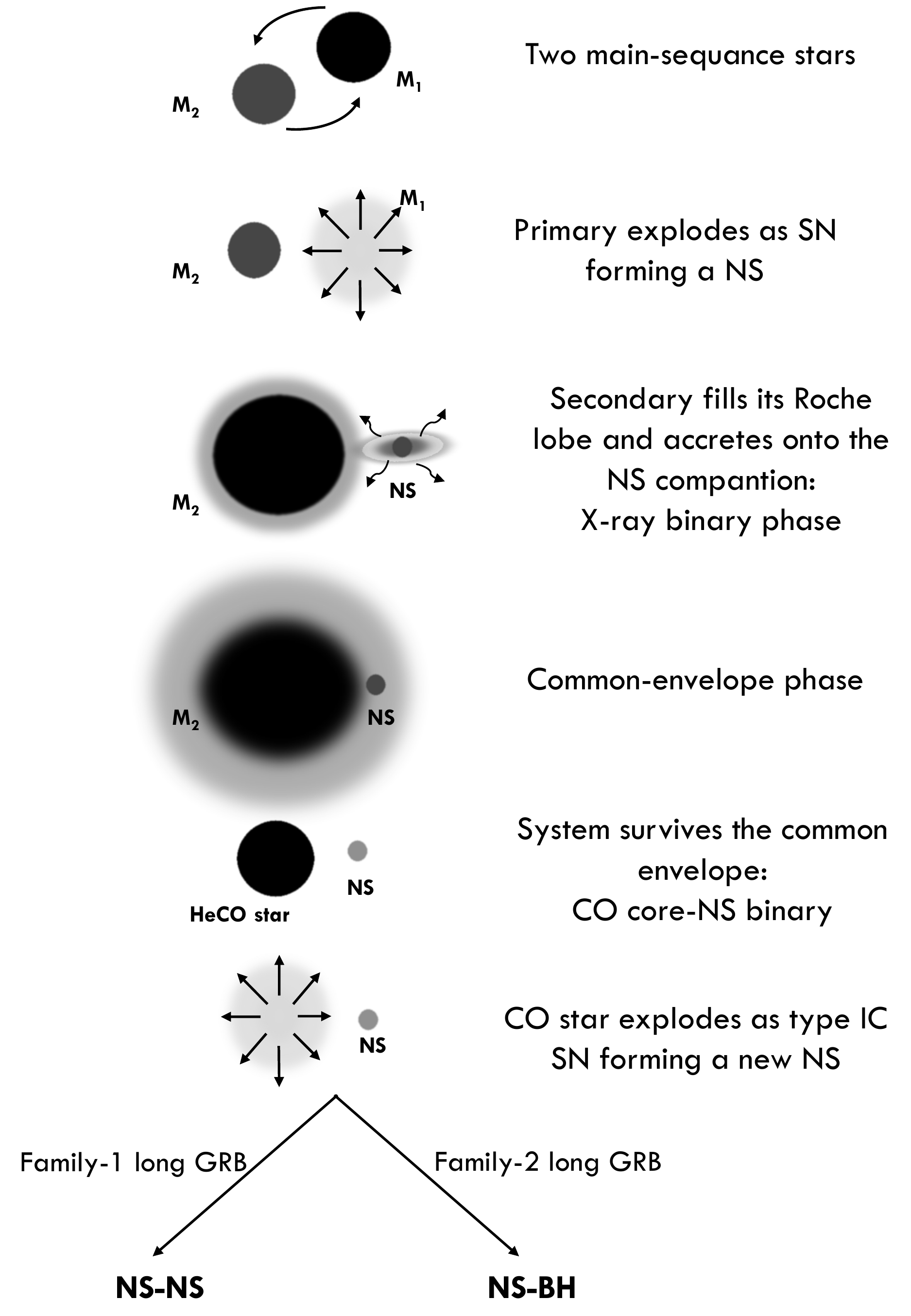}
\caption{Scheme of a possible evolutionary scenario leading to BdHNe as outlined in \citep{2012ApJ...758L...7R}.}\label{fig:evolution}
\end{figure}

The above evolutionary path advanced by \cite{2012ApJ...758L...7R}, see figure~\ref{fig:evolution}, is in agreement with the recent results of \cite{2015MNRAS.451.2123T}, who performed simulations of the evolution a helium core in a tight binary with a NS companion. The helium envelope is stripped-off during the evolution as a result of both mass-loss and binary interactions and at the end it might lead to a SN of type Ib or Ic in presence of the NS companion. Their simulations show the possibility having binaries with orbital periods as short as $\sim 50$~min at the moment of the SN explosion. However, they were interested in the so-called ultra-stripped SNe and therefore they explored systems with helium stars of low initial masses $M_{\rm He}=2.5$--$3.0~M_\odot$, less massive than the ones we expect for the CO cores in BdHNe.

Besides comparing the value of the pre-SN core mass, it is also instructive to compare the radii we have assumed for the pre-SN CO core with the ones obtained by \cite{2015MNRAS.451.2123T}. For example, they obtained a radius of the metal core of the $\approx 3~M_{\odot}$ helium star forming initially a binary of orbital period $0.1$~d with a NS companion of $1.35~M_\odot$, is $R\approx 0.024~R_\odot$. The most similar case we can compare with, corresponds to the CO core formed by the $M_{\rm ZAMS}=15~M_\odot$ progenitor (see table~\ref{tb:ProgenitorSN}), $M_{\rm CO}\approx 3.5~M_\odot$, for which we have adopted a radius of $\approx 4.5\times 10^{9}$~cm$\approx 0.06~R_\odot$. This radius is $\approx 2.5$ times larger than the above $3~M_{\odot}$ helium star of \cite{2015MNRAS.451.2123T}. This implies that our assumption in \citep{2014ApJ...793L..36F} that, due to the 3–-4 orders of magnitude of pressure jump between the CO core and helium layer, the star will not expand significantly when the helium layer is removed, seems to be appropriate. As we discussed in \citep{2014ApJ...793L..36F}, differences of $\sim 2$ in the value of the radius could be due to the different binary interaction ingredients as well as to subtleties of the numerical codes, e.g. between the MESA and the KEPLER codes.

On the other hand, the relatively long possible orbital periods we have obtained (with respect to our minimum value $P_0$) to have BH formation weakens the role of the precise value of the CO core radius on the accretion process and the final fate of the system. If the radius of the CO core is $N$ times the radius of the core of the system with $P=P_0$, then the evolution of the system will be approximately the one of a system with orbital period $P=N^{3/2} P_0$. For example, we have adopted a radius $R_{\rm CO}\approx 7.7\times 10^9$~cm$\approx 0.1~R_\odot$ for the CO core with mass $M_{\rm CO}\approx 9.4~M_\odot$ (see table~\ref{tb:ProgenitorSN}), and thus a minimum orbital period (to have no Roche lobe overflow) of $P_0\approx 5$~min, if it forms a binary with a $2~M_\odot$ NS companion. The maximum value of the orbital period for which we obtained BH formation for those mass parameters was $P_{\rm max}\approx 73$~min (for the NL3 EOS), which would imply that even with a CO core $\approx 15^{2/3}\approx 6$ times larger, BH formation would occur. Despite this fact, the precise value of the CO core mass and radius depends on the binary interactions, hence on the evolutionary path followed by the system; therefore it is appropriate to compute the binary evolution proposed in this work to confirm or improve our estimates for the CO core masses and radii.

\section{Conclusions}\label{sec:6}

We have first computed the angular momentum transported by the supernova ejecta expanding in the tight CO core-NS binary system progenitor of BdHNe. We have shown that the angular momentum of the ejecta is high enough to circularize, although for a short time, around the NS forming a kind of thick disk. We have then simulated the process of accretion onto the NS of the supernova ejecta falling into the Bondi-Hoyle region and forming the disk-like structure around the NS. We have computed both the evolution of the NS mass and angular momentum assuming that the material falls onto the NS surface from an inner disk radius position given by the mostly bound circular orbit around the NS. The properties of the mostly bound orbit, namely binding energy and angular momentum, have been computed in the axially symmetric case with full rotational effects in general relativity. We have computed the changes of these properties in a dynamical fashion accounting for the change of the exterior gravitational field due to the increasing NS mass and angular momentum.

In figure~\ref{fig:vfield} we have shown a set of snapshots of our simulation of the velocity field of the supernova ejecta falling into the NS. It is evident from the evolution the increase of size of the Bondi-Hoyle gravitational capture region of the NS with time due to the increase of the NS mass. It becomes also manifestly evident why, as shown by \cite{FRR2015}, on the contrary to the canonical supernova explosion occurring in binary progenitors studied in population synthesis calculations, BdHNe remain bound even if a large fraction of the binary system's mass is lost in the explosion. Indeed, they might exceed the canonical 50\% limit of mass loss without being disrupted. This impressive result is the combined result of: 1) the hypercritical accretion onto the NS companion which alters both the mass and momentum of the binary; 2) the explosion timescale which is on par with the orbital period, hence the mass ejection cannot be assumed to be instantaneous; 3) the bow shock created, as the accreting NS plows through the supernova ejecta, transfers angular momentum acting as a break on the orbit.

We have shown that the fate of the NS depends on its initial mass: for example, we have seen how a NS with an initial mass $M_{\rm NS}(0)=2.0~M_\odot$ reaches the mass-shedding limit, while for higher initial masses, e.g.~2.5~$M_\odot$, the NS ends at the secular axisymmetric instability limit (see figure~\ref{fig:NSa}). Only those NSs reaching the mass-shedding limit are spun-up up to the maximum value of the angular momentum, $J_{\rm NS,max}\approx 0.7 G (M^{J\neq 0}_{\rm crit})^2/c$ (see figure~\ref{fig:NSb}). Since the NS dimensionless angular momentum reaches a maximum value $<1$, the new black hole formed out of the gravitational collapse of the NS is initially not maximally rotating. Further accretion of mass and angular momentum from material kept bound into the system after the BdHNe process might lead the black hole to approach maximal rotation, however it is out of the scope of this work to explore such a possibility and will be the subject of a forthcoming publication. 

\begin{figure}[!hbtp]
\includegraphics[width=\hsize,clip]{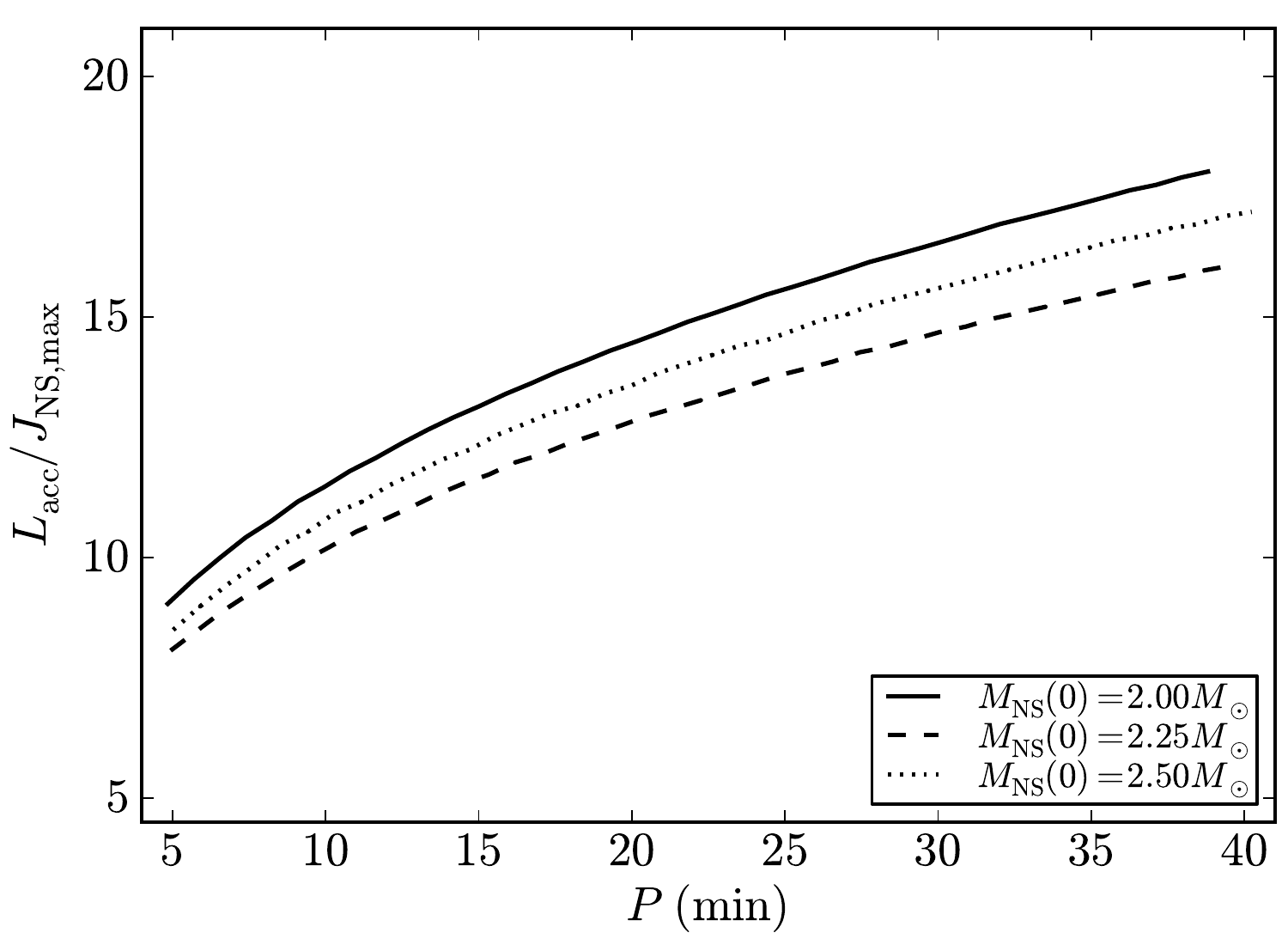}
\caption{Ratio between the total angular momentum transported by the ejecta circularized around the NS and the maximum angular momentum of the NS when it reaches the instability point, for the NL3 EOS.}\label{fig:AngularMomentum_blackhole}
\end{figure}

We can compute the total angular momentum transported by the supernova ejecta material, $L_{\rm acc}$, by integrating the angular momentum per unit time $\dot{L}_{\rm acc}$ during the entire time interval of the accretion process up to the point when the NS reaches the instability region. If we compare the value of $L_{\rm acc}$ with the maximum possible angular momentum that a NS can support, $J_{\rm NS,max}$, we reach a most important conclusion of this work: since $L_{\rm acc} > J_{\rm NS,max}$, see figure \ref{fig:AngularMomentum_blackhole}, there is an excess of angular momentum in the system which might give raise to jetted emission in BdHNe with possible high-energy signatures. 

We have explored the fate of the system for increasing orbital period and as a function of the initial NS mass, for a fixed typical CO core mass. We have shown that, as expected, there exists a maximum orbital period, $P_{\rm max}$, over which the NS does not accrete enough mass to become a BH. Despite that this could be seen as an expected result, it has been quite unexpected, as shown in figure~\ref{fig:Pmax}, that the orbital period does not need to be as short as $\sim 5$~min in order to produce the collapse of the NS to a BH. Indeed, for $M_{\rm NS}(0)=2~M_\odot$ the orbital period can be as longer as $73$~min, for the NL3 EOS. Clearly, the value of $P_{\rm max}$ increases for lower values of the critical mass of the NS, i.e. the softer the EOS the longer the maximum orbital period up to which BH formation occurs, and viceversa. As discussed by \cite{2015ApJ...798...10R}, the existence of $P_{\rm max}$ and the precise value of the NS critical mass define a dichotomy in the long GRBs associated to SNe (BdHNe), namely two long GRB families with different observational signatures: Family-1 long GRBs in which a BH is not produced, and Family-2 long GRBs in which a BH is produced. The relative rate of Family-2 long GRBs with respect to Family-1 long GRBs can give us crucial information on the value of the NS critical mass, hence on the stiffness of the nuclear EOS; and thus population synthesis analyses leading to theoretically inferred rates of events are needed to unveil this important information \citep[see, e.g.,][for the complementary case of short GRBs]{2014arXiv1412.1018R,2015arXiv150407605F}. As a first starting point toward such an analysis, we have discussed a possible evolutionary scenario leading to tight CO core-NS binaries (see figure~\ref{fig:evolution}).

To conclude, in this work we have advanced the first estimates of the role of the angular momentum transported by the supernova ejecta into the final fate of the NS companion in a BdHNe. In order to keep the problem tractable and to extract the main features of these complex systems we have adopted a series of approximations. 1) We have applied the Bondi-Hoyle-Lyttleton formalism to compute the accretion rate onto the NS; BdHNe are time-varying systems which might challenge the validity of this framework which is valid for steady-state systems \citep[see][and references therein]{2004NewAR..48..843E}. 2) We have adopted Taylor series expansions of the supernova ejecta density and velocity around the NS under the assumption that the Bondi-Hoyle radius is small as compared with the binary separation; this has to be considered a first order solution to the problem since for the conditions of BdHNe we have $R_{\rm cap}\sim (0.01$--$0.1)~a$. 3) We have adopted an homologous expansion model for the supernova ejecta which could lead to the suspicion of producing artificially higher accretion rates onto the NS due to the low-velocity inner layers of the ejecta. However, we have already shown \citep[see figure 3 in][]{2014ApJ...793L..36F} that the homologous model leads to an accretion process lasting for a longer time with respect to more realistic explosion models but with lower accretion rates such that the time-integrated accretion rate leads to a lower amount of accreted material by the NS. The reason for this is that, in a given time interval, some of the low-velocity ejecta do not have enough time to reach the gravitational capture region of the NS. 4) We have adopted some characteristic values for the homologous expansion parameter of the supernova ejecta ($n=1$), for the initial velocity of the outermost supernova ejecta layer ($v_{0_{\rm star}}=2\times 10^9$~cm~s$^{-1}$), for the efficiency of the angular momentum transfer from the circularized matter to the NS ($\xi=0.5$). Thus, a systematic analysis of simulations exploring the entire possible range of the above parameters as well as full 2D and/or 3D of the supernova explosion and accretion in BdHNe are required in order to validate and/or improve our results.
 
\acknowledgements{J.~A.~R. acknowledges the support by the International Cooperation Program CAPES-ICRANet financed by CAPES, Brazilian Federal Agency for Support and Evaluation of Graduate Education within the Ministry of Education of Brazil.}


\appendix

\section{Neutron Star Structure}\label{app:NSinfo}

The contents of this appendix are based on the recent work of \cite{2015Cippolletta}. The interior and exterior metric of a uniformly rotating NS can be written in form of the stationary axisymmetric spacetime metric
\begin{equation}
\label{eq1}
ds^2 = -e^{2 \nu} dt^2 + e^{2 \psi} (d \phi - \omega dt)^2 + e^{2 \lambda} (dr^2 + r^2 d \theta^2),
\end{equation}
where $\nu$, $\psi$, $\omega$ and $\lambda$ depend only on variables $r$ and $\theta$. It is useful to introduce the variable $e^{\psi} =  r \sin(\theta) B e^{-\nu}$, being again $B = B(r,\theta)$. The energy-momentum tensor of the NS interior is given by
\begin{equation}
\label{eq2}
T^{\alpha \beta} = (\varepsilon + P) u^\alpha u^\beta + P g^{\alpha \beta},
\end{equation}
where $\varepsilon$ and $P$ denote the energy density and pressure of the fluid, and $u^\alpha$ is the fluid 4-velocity. Thus, with the metric given by equation~(\ref{eq1}) and the energy-momentum tensor given by equation~(\ref{eq2}), one can write the field equations as (setting $\zeta=\lambda + \nu$):


\begin{eqnarray}
\label{eqA1a}
&&\nabla\cdot\left( B \nabla \nu  \right) = \frac{1}{2} r^2 \sin^2 \theta B^3 e^{-4\nu} \nabla \omega \cdot \nabla \omega
+ 4 \pi B e^{2\zeta - 2\nu} \left[ \frac{(\varepsilon+P)(1+v^2)}{1-v^2} + 2P \right],\\
\label{eqA1b}
&&\nabla \cdot \left(  r^2 \sin^2 \theta B^3 e^{-4\nu} \nabla \omega \right) = -16 \pi r \sin \theta B^2 e^{2\zeta - 4\nu} \frac{(\varepsilon +P)v}{1-v^2},
\\
\label{eqA1c}
&&\nabla \cdot \left( r \sin(\theta) \nabla B \right) = 16 \pi r \sin \theta B e^{2\zeta - 2\nu} P,
\end{eqnarray}

\begin{eqnarray}
\label{eqA1d}
{\zeta}_{,\mu} =& - & {\left\lbrace \left( 1-{\mu}^2 \right) {\left( 1+r \frac{B_{,r}}{B} \right)}^2 + {\left[ \mu - \left( 1-{\mu}^2 \right) \frac{B_{,r}}{B} \right]}^2 \right\rbrace}^{-1}  \Biggl[\frac{1}{2} B^{-1} \left\lbrace r^2 B_{,rr} - {\left[ \left( 1 -{\mu}^2 \right) B_{,\mu} 
                       \right]}_{,\mu} - 2\mu B_{,\mu} \right\rbrace  \nonumber \\
                       & \times & \left\lbrace - \mu + \left( 1-{\mu}^2 \right) \frac{B_{,\mu}}{B} \right\rbrace +  r\frac{B_{,r}}{B} \left[ \frac{1}{2} \mu + \mu r\frac{B_{,r}}{B} + \frac{1}{2} \left( 1-{\mu}^2 \right) \frac{B_{,\mu}}{B} \right] + \frac{3}{2} \frac{B_{,\mu}}{B} \left[ -{\mu}^2 + \mu \left( 1-{\mu}^2 \right) \frac{B_{,\mu}}{B} \right]\nonumber\\
                       & - & \left( 1-{\mu}^2 \right) r\frac{B_{,\mu r}}{B} \left( 1+r\frac{B_{,r}}{B} \right) - \mu r^2 {\left( {\nu}_{,r} \right)}^2 - 2\left( 1-{\mu}^2 \right) r{\nu}_{,\mu}{\nu}_{,r} + \mu \left( 1-{\mu}^2 \right){\left( {\nu}_{,\mu} \right)}^2 - 2\left( 1-{\mu}^2 \right) r^2 B^{-1} B_{,r} {\nu}_{,\mu} {\nu}_{,r}\nonumber \\
                       & + & \left( 1-{\mu}^2 \right) B^{-1} B_{,\mu} \left[ r^2 {\left( {\nu}_{,r} \right)}^2 - \left( 1-{\mu}^2 \right) {\left( {\nu}_{, \mu} \right)}^2 \right] + \left( 1-{\mu}^2 \right) B^2 e^{- 4\nu} \Bigl\lbrace \frac{1}{4} \mu r^4 {\left( {\omega}_{,r} \right)}^2 + \frac{1}{2} \left( 1-{\mu}^2 \right) r^3 {\omega}_{,\mu} {\omega}_{,r}\nonumber \\
                       & - & \frac{1}{4} \mu \left( 1-{\mu}^2 \right) r^2 {\left( {\omega}_{, \mu} \right)}^2 + \frac{1}{2} \left( 1-{\mu}^2 \right) r^4 B^{-1} B_{,r} {\omega}_{,\mu} {\omega}_{,r} - \frac{1}{4} \left( 1-{\mu}^2 \right) r^2B^{-1}B_{,\mu} \left[ r^2 {\left( {\omega}_{,r} \right)}^2 - \left( \-{\mu}^2 \right) {\left( {\omega}_{,\mu} \right)}^2 \right] \Bigr\rbrace \Biggr],
\end{eqnarray}

%
where, in the equation for ${\zeta}_{,\mu}$, we introduced $\mu\equiv \cos(\theta)$. 

We can integrate numerically the above Einstein equations once a relation between $\varepsilon$ and $P$ is given,, namely an equation of state (EOS). The NS interior is made of a core and a crust. The core of the star has densities higher than the nuclear value, $\rho_{\rm nuc}\approx 3\times 10^{14}$~g~cm$^{-3}$, and it is composed by a degenerate gas of baryons (e.g.~neutrons, protons, hyperons) and leptons (e.g.~electrons and muons). The crust, in its outer region ($\rho \leq \rho_{\rm drip}\approx 4.3\times 10^{11}$~g~cm$^{-3}$), is composed of ions and electrons, and in the so-called inner crust ($\rho_{\rm drip}<\rho<\rho_{\rm nuc}$), there are also free neutrons that drip out from the nuclei. For the crust, we adopt the Baym-Pethick-Sutherland (BPS) EOS \citep{1971ApJ...170..299B}. For the core, we here adopt modern models based on relativistic mean-field (RMF) theory, which have Lorentz covariance, intrinsic inclusion of spin, a simple mechanism of saturation for nuclear matter, and they do not violate causality. We use an extension of the formulation of \cite{boguta77} with a massive scalar meson (sigma) and two vector meson (omega and rho) mediators, and possible interactions between them. 

With the knowledge of the EOS we can compute equilibrium configurations integrating the above equations for suitable initial conditions, for instance central density and angular momentum (or angular velocity) of the star. Then, after integrating the Einstein equations, properties of the NS can be obtained for the given central density and angular momentum such as the total gravitational mass, the total baryon mass, polar and equatorial radii, moment of inertia, quadrupole moment, etc. 

For the present problem of accretion onto the NS, as we have mentioned in section~\ref{sec:2}, an important relation to be obtained from the NS equilibrium properties, is the one between the total baryon rest-mass, $M_b$, and the gravitational mass, $M_{\rm NS}$, namely the gravitational binding energy of the NS; see equation (\ref{eq:mgmb}). For non-rotating configurations, \cite{2015Cippolletta} obtained an EOS-independent relation
\begin{equation}\label{eq:mbm}
\frac{M_b}{M_\odot}=\frac{M_{\rm NS}}{M_\odot}+\frac{13}{200}\left(\frac{M_{\rm NS}}{M_\odot}\right)^2.
\end{equation}

For the non-zero angular momentum configurations we are interested at, \cite{2015Cippolletta} obtained
\begin{equation}
\frac{M_b}{M_\odot}=\frac{M_{\rm NS}}{M_\odot}+\frac{13}{200}\left(\frac{M_{\rm NS}}{M_\odot}\right)^2\left(1 - \frac{1}{130} j_{\rm NS}^{1.7}\right),
\end{equation}
where $j_{\rm NS}\equiv c J_{\rm NS}/(G M^2_\odot)$. This formula is accurate within an error of 2\% and it correctly generalizes the above equation~(\ref{eq:mbm}), aproaching it in the limit $j_{\rm NS}\to 0$.

The NS can accrete mass until it reaches a region of instability. There are two main instability limits for rotating NSs, namely the mass-shedding or Keplerian limit, and the secular axisymmetric instability. \cite{2015Cippolletta} have shown that the critical NS mass along the secular instability line, is approximately given by:
\begin{equation}
M_{\rm NS}^{\rm crit}=M_{\rm NS}^{J=0}(1 + k j_{\rm NS}^p),
\end{equation}
where the parameters $k$ and $p$ depends of the nuclear EOS (see table \ref{tb:StaticRotatingNS}). These formulas fit the numerical results with a maximum error of 0.45\%.
\begin{table}[!hbtp]
\centering
\caption{Critical mass and corresponding radius for selected parameterizations of nuclear EOS}\label{tb:StaticRotatingNS}
\begin{tabular}{cccccccc}
\hline \hline
EOS  &  $M_{\rm crit}^{J=0}$~$(M_{\odot})$ & $R_{\rm crit}^{J=0}$~(km) & $M_{\rm max}^{J\neq 0}$~$(M_{\odot})$ & $R_{\rm max}^{J\neq 0}$~(km) &$p$&$k$ & $f_{K}$~(kHz) \\ 
\hline
NL3 & $2.81$&$13.49$ &$3.38$ & 17.35 & $1.68$&$0.006$ & $1.34$\\
GM1 & $2.39$&$12.56$ &$2.84$& 16.12&$1.69$&$0.011$ & $1.49$\\
TM1 & $2.20$ &$12.07$ &$2.62$ & 15.98 &$1.61$&$0.017$ & $1.40$\\  
\hline
\end{tabular}
\tablecomments{In the last column we have also reported the rotation frequency of the critical mass configuration in the rotating case. This value corresponds to the frequency of the last configuration along the secular axisymmetric instability line, i.e the configuration that intersects the Keplerian mass-shedding sequence.}
\tablerefs{\cite{2015Cippolletta}.}
\end{table}

In addition to the above relations, we have used in this work an analytic formula, equation (\ref{eq:lisco}) obtained by \cite{2015Cippollettainprep}, which gives us the angular momentum of the mostly bound circular orbit around a uniformly rotating NS, as a function of the NS mass and angular momentum. Such a relation has allowed us to perform the simulations of the evolution of the accreting NS in a semi-analytic fashion, including dynamically the feedback of the increase of the NS mass and angular momentum into the exterior geometry.

\end{document}